\documentclass[11pt]{article}
\usepackage{a4}
\usepackage{amssymb}
\usepackage{rotating}
\usepackage{array}
\usepackage{hyperref}

\let\pa=\partial
\def\su{\ifmmode SU(2) \else $SU(2)$\fi}

\begin{document}
\setlength{\extrarowheight}{2pt}

\title{Elements of dodecahedral cosmology.}
\author{Marc P. Bellon\\
\small  Laboratoire de Physique Th\'eorique et Hautes Energies\thanks{Tour 24-25, 5\` 
eme \'etage, Boite 126, 4 Place Jussieu,  75252 Paris Cedex 05. }, 
(LPTHE)\\
\small Unit\'e Mixte de Recherche UMR 7589\\
\small Universit\'e Pierre et Marie Curie-Paris6; CNRS; Universit\'e Denis 
Diderot-Paris7}
\date{}
\maketitle
\begin{abstract}
A general method for the determination of the harmonics of  quotients of the 3-sphere 
is given.  They can all be deduced from three objects already known from Klein. We 
further show explicitly how these harmonics can be organized in irreducible 
representations of the holonomy group: this allows for the determination of 
the full correlation matrix of the Cosmic Microwave Background in the spherical 
harmonics basis. 

\end{abstract}

\section{Introduction}
The question of the global structure of our universe is a basic question in 
cosmology. Inflationary models of the universe suppose that the expansion is so great that the 
observable part of the universe will for ever remain negligible and global aspects 
firmly out of the reach of observation. However the low values of 
the quadrupole and octopole of the cosmological microwave background observed by 
WMAP~\cite{WMAP1} suggested that the universe may not be so large  and a case has been 
made for its structure to be that of a dodecahedral quotient of the three-sphere 
in~\cite{luminet}. These low values are by themselves no 
proof of such a finite universe, since other explanations have been proposed,  the 
simplest one being that these low values are compatible with cosmic variance.  It is 
therefore important to have a more compelling argument for a small compact universe, 
which also allows to choose between the different proposed 
geometries. 

The purpose of this paper is to show that a general method to describe the harmonics 
can be deduced from the classical results of Klein~\cite{klein} on the invariant 
polynomials for subgroups of $SU(2)$. 
Furthermore the correlations between different multipoles become computable, when 
using the decomposition of the representation of \su\ in irreducible representations of 
the binary polyhedral group. The count of these representations is known from the 
study of Kostant~\cite{Ko,Ko2}, but explicit basis for these 
representations are necessary and will be presented here. Their constructions owe 
much to the methods of Klein.

In the study of the cosmological microwave background in a given geometry, the 
eigenmodes of the Laplacian on this space have to be determined. In the case of the 
quotients of the three-sphere, the eigenvalues with their multiplicities have been 
determined by considerations on the characters of the binary polyhedral 
groups on  the space of spherical harmonics~\cite{Ik}. The determination of the 
eigenfunctions have been less straightforward. In the first studies, numerically demanding 
image methods have been used, which limited the investigation to a small number of 
eigenmodes~\cite{luminet}. Even if simplifications have 
been done in the calculations, the methods presented in previous works do not allow 
to generate harmonics of arbitrary degree. This problem has been alleviated when 
computing only the spherical averages of the harmonics by the remark made 
in~\cite{AuLuSt05} and first proved in~\cite{Gu05} that the number of modes of a 
given degree is sufficient in this case. It should be noted that the general 
construction of the eigenmodes has been given in~\cite{We05}, but failing to relate them to a 
basis more appropriate for the evaluation of the fluctuations of the microwave 
background. The two basis with 
their mutual relations were presented in~\cite{LRC05}, however with a level by level 
approach to the determination of the invariants.

In the following, I will first review how the peculiar nature of the four dimensional 
rotation group $SO(4)$, which is essentially the product of two copies of the group 
\su, gives a basis for the functions on the three dimensional sphere, where the group 
$\Gamma$ through which we factor has a simple expression. From 
the point of view of a given observer, an other basis with simple transformations 
through the rotations around this observer is more convenient.
These two basis are related by the Clebsch--Gordan coefficients of \su.

The functions on the quotients of the three-sphere are then described and arranged in 
irreducible representations of the finite subgroup $\Gamma$ of the group \su\ 
through which we quotient. This allows to predict a simple form for the correlations 
of the microwave background in a suitable basis.

Finally, the building blocks for the invariants and the other irreducible 
representations of the finite subgroups of \su\ are given explicitly.  This study extensively 
uses the description of the irreducible representations of \su\ by homogeneous 
polynomials in two variables. This allows for easy computations of the 
actions of \su\ by substitutions and the generation of new objects of interest by 
multiplication and other elementary operations on polynomials.

\section{Functions on the three-sphere.}
The three-sphere is remarkable among spheres because it can be given the structure of 
a non-Abelian group. This group can be considered either as the one of the 
quaternions of  unit norm or the group of unitary two by two matrices $SU(2)$. As a 
group manifold, the three-sphere has symmetries stemming from the left  and 
right actions on itself, which are different since the group is non-Abelian. The 
rotation group in four dimensions, the symmetry group of the three-sphere, is therefore 
factorized in a pair of factor homeomorphic to $SU(2)$, which will be denoted 
$SU(2)_{L}$ and $SU(2)_{R}$.
Functions on the three-sphere form representations of this product group.

In order to make the description, it will be useful to parameterize the sphere with 
complex coordinates, stemming from the general form of matrices in the group 
$SU(2)$:
\begin{equation}
	\label{su2}  p = \pmatrix{s&-\bar t\cr t& \bar s\cr}
\end{equation}
The two complex numbers $s$ and $t$ verify the constraint $s \bar s + t \bar t=1$. 
Unconstrained real coordinates for the 3-sphere can be taken as the phases of the 
complex variables $s$ and $t$, $\psi$ and $\phi$, each varying in the interval 
$[0,2\pi]$ and a third angle $\tau$, varying in $[0,\pi/2]$ to denote their 
relative modules.
\begin{equation}
	s = \cos \tau \; e^{i\psi},\quad t = \sin \tau \; e^{i\phi}.
\end{equation}
These angles are linked to the parameterization of three dimensional rotations by the 
Euler angles, which can be taken as $\psi + \phi$, $2\tau$ and $\psi-\phi$. The 
metric in these coordinates take the simple form $ dl^{2} = d\tau^{2} + \cos^{2}\tau 
d\psi^{2} + \sin^{2}\tau \,d\phi^{2} $ and the volume element is 
$\sin\tau\cos\tau \,d\tau \,d\psi \,d\phi$.

From the matrix form~(\ref{su2}), it appears that its four complex components form 
two doublets of the $SU(2)_{L}$ group, $(t,-s)$ and $(\bar s, \bar t)$, and two of the 
$\su_{R}$ group, $(s,-\bar t)$ and $(t,\bar s)$. They are in the tensor product of 
the fundamental representations of the two \su\ groups. When taking 
polynomials in the basic variables $s$ and $t$ and their complex conjugates, one 
obtains only product of identical representations of the left and right \su\ groups. One 
usually denotes them by a pair $(j,j)$ of the spin with respect to each of the \su\ 
factors. Acting with the two commuting operators 
$t\partial_{s}-\bar s \partial_{\bar t}$ and $-\bar t \partial_{s}+\bar s 
\partial_{t}$ on $s^{2j}$, one obtains $(2j+1)^{2}$ harmonic polynomials of degree $2j$ which 
form a basis of the space of eigenvalue $2j(2j+2)$ of the Laplacian on the sphere. 
They can be characterized by the pairs $(p,q)$ of the eigenvalues of 
the $L_{z}$ operators in the two groups. Since $j$ is a half integer, $n=2j$ can take 
any integer value. 
In the variables $\tau$, $\psi$, $\phi$, the unnormalized eigenfunctions take a 
rather simple form:
\begin{eqnarray}
     Z_{npq}(\tau,\psi,\phi)& =& (-1)^{\max(j+p,j-q)} R_{npq}(\tau) e^{i(q-p)\psi} 
e^{i(p+q)\phi}\\
     R_{npq}(\tau)  &=& \cos^{|p-q|}\tau \sin^{|p+q|}\tau 
     	P^{|p-q|,|p+q|}_{{1\over2}(n-|p+q|-|p-q|)}(\cos 2\tau)
\end{eqnarray}
The $P^{\alpha,\beta}_{r}$ are polynomials of degree $r$, orthogonal with respect to 
the weight $(1-x)^{\alpha}(1+x)^{\beta}$, known as Jacobi polynomials. Since $p$ and 
$q$ are either both integers or both half-integers, $Z$ is always $2\pi$ periodic in 
$\phi$ and $\psi$.

If the basis of eigenvectors of $L^{z}_{L}$ and $L^{z}_{R}$ is perfect for the 
intrinsic study of the harmonics, from the point of view of an observer, it is more 
convenient to make explicit the representations of the rotations around the observer. 
In the case of the three-sphere, these rotations form the diagonal 
subgroup of the full symmetry group, with the two \su\ factors identified. By the 
standard composition of the representations of \su, the harmonics of degree $n$ 
decompose into the representations with $l=0$, 1,\dots up to $n$. Vectors are 
characterized by the pairs $(l,m)$. The change of basis can be made 
explicit with the Clebsch--Gordan coefficients:
\begin{equation}\label{basechange}
\vert n,l,m\rangle =   \sum_{m_{L},m_{R}} (j  j m_{L} m_{R}\vert l m) \vert 
n,m_{L},m_{R}\rangle 
\end{equation}
The coordinates in which those functions take simple forms are the $(\chi,\theta,\phi)$ variables, where $\chi$ is a measure of the distance to the observer and 
$(\theta,\phi)$ the usual polar coordinates on the 2-sphere. In the change of 
variable from $(\tau,\psi,\phi)$ to $(\chi,\theta,\phi)$, $\chi$ and $\theta$ 
only depend on $\tau$ and $\psi$ and $\phi$ goes to  $\phi - \pi/4 $. The precise 
relation could be obtained from the following expression of the variables $s$ and $t$ in 
terms of $(\chi,\theta,\phi)$:
\begin{equation}
	s = \cos \chi + i \sin \chi \cos \theta,\quad t = i \sin \chi \sin  \theta \; 
e^{i\phi}.
\end{equation}
The factor $i$, which gives a $\pi/4$ difference between the coordinates $\phi$ in 
the two systems of coordinates, allows to be compatible with the usual conventions. The 
$\chi$ dependences of the corresponding functions are expressed in terms of  
Gegenbauer polynomials of degree $n-l$ and do not vary with $m$. The 
Gegenbauer or ultraspherical polynomials are a special case of the Jacobi 
polynomials.
\begin{equation}
	W_{nlm} (\chi,\theta,\phi)= i^{l} \sin^{l}\chi P^{l+{1\over2},l+{1\over2}}_{n-l}(\cos \chi) Y_{lm}(\theta,\phi)
\end{equation}
The factor $i^{l}$ is necessary to be compatible with real Clebsch--Gordan 
coefficients according to the conventions of~\cite{LanLif} that we adopt. The functions 
$W_{nlm}$ are determined by their transformation property under the rotation group 
and are related to the functions $Z_{npq}$ by eq.~(\ref{basechange}). The 
only difficulty is to verify the phase, but it can be determined by consideration of 
the leading terms of some special cases. 

The linear contribution of a perturbation of the universe by such a mode to the 
cosmological background is simple: rays arriving at the observer are characterized by 
fixed $\theta$ and $\phi$, so that the angular dependence is simply given by the 
$Y_{lm}$ part. For the Sachs--Wolfe effect in the instantaneous 
approximation, the value of the $\chi$ dependent part at the value $\chi_{LS} $ 
corresponding to the surface of last scattering will appear. The details of the derivation 
and the computation of other contributions can be found elsewhere (see 
e.g.,~\cite{AuLuSt05}), but what is important is that the time variation of 
the perturbations only depend on $n$ and the radial one only on $n$ and $l$: the 
modes which differ only by the value of $m$ appear with the same weight.

\section{Quotients of the three-sphere.}
The definition of the Poincar\'e space and other cosmologically interesting quotients 
of the 3-sphere relies on the special structure of the four-dimensional rotation 
group as a direct product of two factors. Each of its \su\ factors acts without fixed 
points: any finite subgroup $\Gamma$ of one of these factors also 
acts without fixed points so that the corresponding quotients of the three-sphere are 
regular manifolds. All the possible regular quotients of the three-sphere are not of 
this form, since  double action manifolds appear in the classification done in the 
year 1932 in~\cite{TS30,TS32}. Functions on the quotients 
are in one to one correspondence with functions on the three-sphere invariant under 
these finite subgroups.

Viewed as elements of the tensor product of representations of the two \su\ factors 
of $SO(4)$, the definition of invariants is simple: they are tensor product of an 
invariant of the group $\Gamma$ and an arbitrary vector. Modes of the Laplacian with 
eigenvalue $n(n+2)$ comes therefore in groups of $n+1$ since 
$n=2j$.

The problem of the determination of the eigenmodes of the Laplacian for such a space 
therefore reduces to the determination of the invariants of the subgroup $\Gamma$ in 
the representations of \su.

However the knowledge of the invariants is not necessary if one is only interested in 
the spherically averaged fluctuations of the microwave background, their number in 
each representation of \su\ is sufficient. This stems from a sum rules for the 
coefficients of the eigenmodes first inferred from numerical evidence 
in~\cite{AuLuSt05} and proposed as a conjecture in their equation~(25):
\begin{equation}\label{sumrule}
	\sum_{i}\sum_{m=-l}^{l} \left| C^{i}_{nlm}\right|^{2} = J({n}){2l+1\over 
n+1}
\end{equation}
Here, $i$ indexes an orthonormal basis of the modes of the quotient of the sphere at 
level $n$ and the $C^{i}_{nlm}$ are the components of these modes in the basis of the 
$W_{nlm}$ and $J(n)$ is the number of invariants of the group $\Gamma$. The right 
hand side is modified from the one given in~\cite{AuLuSt05} by the 
change to $n=\beta-1$ and the total number of modes is $J(n)$ times $n+1$.
A first proof was given in~\cite{Gu05}. In this new proof, the symmetry of the 
Clebsh--Gordan coefficients expressed through the $3j$-symbols are used to obtain a sum 
rule for their squares. 

We first remark that the relation~(\ref{sumrule}) is simply verified in the case of 
the sphere. The group $\Gamma$ being reduced to the identity, all vectors are 
invariant, $J(n)$ equals $n+1$ and the right hand side is simply $2l+1$. On the other 
hand, we can choose as a basis of the modes the one indexed by $l$ and 
$m$, so that the sum on $m$ in the left hand side is 1 or 0 according if $i$ 
corresponds or not to the given $l$. Since there are $2l+1$ modes for each $l$, we get also 
$2l+1$ in the left hand side.

The fundamental identity which will be used is the following one: 
\begin{equation} \label{fund}
\sum_{m_{1},m_{2}}  \vert ( j j, m_{0} m_{1}\vert l m_{2}) \vert^{2} = { 2 l +1 \over 
2 j +1 }
\end{equation}
If the sum were on $m_{0}$ and $m_{1}$, it would simply be the unitarity of the 
decomposition of the state $\vert l,m_{2}\rangle$ in the base of tensor products of two 
spins and the right hand side would be 1. The use of Wigner's $3j$-symbols allows to 
deduce eq.~(\ref{fund}) from the unitarity condition. The 
$3j$-symbols are invariant by cyclic permutation of the three columns and take at 
most a sign for a permutation of two columns. They allow to express the Clebsch--Gordan 
coefficients as~\cite{LanLif}\footnote{Other choices of normalization are used. This 
one has the advantage of giving rise to real Clebsch--Gordan 
coefficients and uniform treatment of integer and half-integer values of $l$ but 
implies that the eigenfunctions for $l$ odd and $m$ zero are imaginary. This is of no 
import in this work where we only consider absolute values, but a coherent set of 
phase choices must be made.}
\begin{equation}\label{3j}
(j_{1} j_{2}, m_{1} m_{2}\vert l m) = (-1)^{j_{1}-j_{2}+m}\sqrt{2l+1}  \left(\begin{array}{ccc} j_{1} & j_{2}  & l  \\ m_{1} & m_{2}  & -m \end{array}\right) 
\end{equation}
Hence we have:
\begin{equation}\label{bas}
(j j, m_{0} m_{1}\vert l m_2) = \pm \sqrt{2l+1\over 2j+1} ( j m_0 \vert j l , -m_1 
m_2)
\end{equation}
Taking the square of this expression and summing on $m_1$ and $m_2$, we get the 
equation~(\ref{fund}) using the unitarity of the decomposition of the vector $|j 
m_{0}\rangle$.

Suppose now that the invariant vector $v$ of degree $2j$ of the group $\Gamma$ is an 
eigenvector of $L_{z}$, $v= \vert m_{0}\rangle$, for a given $m_{0}$. The invariant 
functions form the family $\vert m_{0},m_{1}\rangle$ with $m_{1}$ playing the role of 
the index $i$. The relation~(\ref{sumrule}) is then simply the 
identity~(\ref{fund}).  A general invariant $v$ is a linear combination of such 
vectors $|m_{0}\rangle$. The left hand side of the identity~(\ref{fund}) is quadratic in 
the coordinates of $v$; the square terms give back the identity~(\ref{fund}) for a 
normalized vector $v$ while the interference terms vanish: 
terms with different $m_{0}$ contribute to $C^{i}_{nlm}$ with differing values of 
$m$.  Finally in the case where there are different invariants of the same degree $j$, 
the modes associated to each of the invariants are uncorrelated, so that their 
contributions simply add. 

Using the relation~(\ref{sumrule}) and the independence on $m$ of the radial parts of 
the eigenmodes of the sphere, one can deduce that the contributions of eigenmodes of 
the Laplacian of degree $2j$ to the total power in degree $l$ of the microwave 
background  is the one of the corresponding modes on the sphere 
multiplied by the number of invariants of $\Gamma$ of this degree divided by $2j+1$, 
up to the normalization of the initial power spectrum.

\section{Correlations.}

The power spectrum is however not sufficient to characterize the geometry. The effect 
of the geometry is important only for the lowest harmonics, which are plagued by 
their large cosmic variance and similar results can be obtained with quotients by the 
dodecahedral or cubic groups.  One therefore want to compute the 
full correlation matrix of the spherical harmonics. For a microwave sky, this matrix 
is a really big object, since the total number of spherical harmonics grows 
quadratically with the maximum $l$ considered, and the correlation matrix is 
quadratic in the number of objects considered. For the  values of $l$ from 
2 to 10, we have 117 different spherical harmonics, which means 6903 coefficients in 
the full correlation matrix. The problem is double: how to compute efficiently all 
these numbers and how to compute the relevance of a given model for our present 
universe. We only have the 117 measured coefficients\footnote{In fact, 
the galactic foreground makes some of these measures impossible, roughly in the 
proportion of the solid angle that the galactic cut  covers on the sky. This only worsen 
the problem.} of one realization to estimate the much bigger number of correlations. 
In the isotropic simply 
connected spaces, the problem simplifies a lot since the different harmonics are 
uncorrelated and the correlation matrix is fully determined by the 9 amplitudes of the 
$\langle C_{l}\rangle$, which can therefore be well constrained from the 
observations, even if the limited number of harmonics with small $l$ limits the 
precision of the determination of the $\langle C_{l}\rangle$. 

In fact, in the quotients of the three-sphere, the homotopy group $\Gamma$ appears 
also as a rotational symmetry around any observer. It is not a symmetry of the 
perturbations, but it is a symmetry of their correlations.
This symmetry allows to characterize the correlation matrix with a much smaller 
number of coefficients, 310 in the octahedral case and 130 in the dodecahedral case. This 
simplification of the correlation matrix is made manifest when decomposing the 
spherical harmonics in irreducible representations of 
$\Gamma$.

At a given degree $n$, the eigenmodes of the space are a tensor product of an 
invariant $v$ of the group $\Gamma$ and a vector in the representation of dimension $n+1$ of 
\su. We take as a basis of this representation of \su\  one adapted to its 
decomposition in irreducible representations of $\Gamma$. Its vectors 
$u_{r,i}$ are indexed by the representation $r$ and the index $i$ in the 
representation. We then have to decompose the tensor product $v\otimes u_{r,i}$ into the 
different representations of \su\ with $l$ varying between 0 and $n$. As $v$ is 
invariant under $\Gamma$, all the terms of this decomposition will 
transform under $\Gamma$ as the vector of index $i$ in a representation of type $r$. 
This mode will therefore contribute to fluctuations of the background which are 
themselves vector of index $i$ in a component $r$ of the decomposition of the 
spherical harmonics in representations of $\Gamma$.

With the usual assumption that the diverse modes of the universe are uncorrelated, we 
deduce that only vectors with the same index in the same representation of $\Gamma$ 
can be correlated. Furthermore, the vectors with different indexes of a given 
representation have the same correlation matrix from the invariance 
under $\Gamma$ of the whole procedure. The procedure for the explicit calculation of 
the correlations will be detailed in a further work.

If we retain the example of limiting ourselves to a maximum $j$ of 10, we have in the 
case of the dodecahedron a total of two trivial representations, excluding the one 
of spin zero, five times the $3$ representation, excluding the unobservable  dipole 
at $l=1$, ten times the $5$ representation, eight times the $4_{v}$ 
and six times the $3'$ representations. The fact that the number of times each 
representation is present is roughly proportional to its dimension is exact asymptotically, 
as can be seen from the formulas of Kostant. The correlation among the fifty vectors 
appearing in $5$ representations is given by five 
times the same symmetric ten by ten matrix, that is 55 coefficients. Adding the 
contribution for the other irreducible representations gives the announced number of 130 
coefficients.  The fact that this simplification of the correlation is manifest in a 
special basis is not a limitation in practice, since a change of 
basis is in any case necessary to rotate the coordinates from the galactic frame to 
the one adapted to the unknown orientation of the group $\Gamma$. 

As we shall see in the following section, the determination of the basis elements can 
be reduced to the expansion of products of polynomials, an easy task for any 
computer algebra system.

\section{Explicit formulas.}\label{McK}
\subsection{Generalities.}

It is time to produce the explicit forms of the invariant vectors and the 
decomposition of the representations of \su\ in irreducible representations of the group 
$\Gamma$.
The question of the invariants has been completely solved by Klein more than a 
century ago~\cite{klein}. It was not however described in these terms, since Klein searched 
for invariant functions on the ordinary sphere under subgroups of the rotation group 
$SO(3)$.  However his study is based on the description of the 
sphere as the complex projective space $\mathbb P^{1}(\mathbb C)$ and the rotation of 
the sphere are represented as two by two matrices of complex numbers. These two by 
two matrices are to be taken projectively, but the unimodularity condition can be 
used to reduce this ambiguity to a sign. Klein remarked that 
in the case of the symmetry groups of regular polyhedrons, this choice of sign cannot 
be made in a way compatible with the group structure, so that one has to consider a 
double cover of the groups and he therefore studied finite subgroups of \su, even if 
he did not use this modern terminology. 

What is remarkable in Klein construction is that a unique invariant is sufficient to 
create the whole family of invariants. The irreducible representations of \su\  are 
symmetric tensor products of the fundamental one, and if we take two variables 
$(s,t)$ that transform as a fundamental representation of \su, all 
irreducible representations can be obtained as homogeneous polynomials in the 
variables $s$ and $t$. Polynomials of total degree $d$ give the $j=d/2$ representation of 
\su. The generators of the Lie algebra of \su\ are written as:
\begin{equation}
L_{z} = {1\over 2} (s \partial_{s}-t\partial_{t}), \quad L_{-} = t \partial_{s}, 
\quad L_{+} = s \partial_{t}
\end{equation}
The monomial $s^{p}t^{q}$ has eigenvalue $(p-q)/2$ under $L_{z}$ and its squared norm 
evaluate to $\pmatrix{p+q \cr p}^{-1}= p! q!/(p+q)! $. The action of the group \su\ 
is simply obtained by linear substitutions on the variables $s$ and $t$. It is then 
evident that the product of invariant polynomials will give a new 
invariant polynomial. But other combinations are possible, based on the use of 
partial derivatives. The doublet $( \pa_{s},\pa_{t})$ transforms as a dual basis to the 
fundamental basis $(s,t)$ and the properties of the invariant antisymmetric tensor 
$\varepsilon$ of \su\ make $(-\pa_{t},\pa_{s})$ transform as 
$(s,t)$. It then results that if the polynomials $P$ and $Q$ are both invariant under 
a subgroup of \su, a new invariant polynomial is given by:
\begin{equation}\label{combeq}
	C(P,Q) = \left | \matrix{ \pa_{s} P & \pa_{s}Q \cr \pa_{t} P & \pa_{t}Q \cr }\right 
| 
		= \pa_{s} P \pa_{t} Q - \pa_{t} P \pa_{s} Q .
\end{equation}
This expression being antisymmetric in $P$ and $Q$, it cannot produce anything non 
trivial from a single invariant, but the Hessian can.
\begin{equation}\label{Hess}
\mathrm{Hess}(P) = \left | \matrix{ \pa_{s}^{2} P & \pa_{s}\pa_{t}P \cr 
\pa_{s}\pa_{t} P & \pa_{t}^{2}P \cr }\right | 
		= \pa_{s}^{2} P \pa_{t}^{2} P - (\pa_{s}\pa_{t} P )^{2} .
\end{equation}
From a first invariant of degree $d$, a second one with degree $2d-4$ is obtained as 
the Hessian, and the combination of these two invariants by $C$ gives a third one of 
degree $3d-6$. This three invariants are sufficient to generate all invariant 
polynomials by multiplication. In fact, these equations are special cases of
the formula of appendix~\ref{comb} for the combination of two representations:
equation (\ref{combeq}) is the case $r=1$ and equation (\ref{Hess})
the case $r=2$ with identical $P$ and $Q$.

This method of Klein can be generalized to the problem of giving not only the 
invariants of $\Gamma$, that is the objects transforming in the trivial one-dimensional 
representation of $\Gamma$, but also all irreducible representations of 
$\Gamma$.
Bertram Kostant made use of the McKay correspondence between finite subgroups of \su\ 
and Lie algebras of $ADE$ type~\cite{McK2,McK3,GV} to count the irreducible 
representations of the binary polyhedral groups in the representations of 
$SU(2)$~\cite{Ko,Ko2,St}. For our purpose it is however necessary to have explicit 
realizations of these representations, but the knowledge of the number of 
representations to find is an important guide. The structure of the generating functions for the 
multiplicities of a given representation is simple: it is a polynomial $P(t)$ divided 
by $(1-t^{2a})(1-t^{2b})$, with $2a$ and $2b$ the 
degrees of the first two invariant polynomials $P$ and $Q$ derived by Klein, which 
have no relation between them. If we take an irreducible representation of the group 
$\Gamma$ realized as homogeneous polynomials of degree $n$, we can produce a new 
realization in degree $n+ 2a k + 2b l $ by multiplying these 
polynomials by  $P^{k}Q^{l}$. The formula derived by Kostant shows that this exhausts 
the occurrences of such a representation in any representation of \su, if we have 
realizations of the representations of the degrees $n$ appearing as exponents in the 
polynomial $P(t)$. Kostant proved that 
the number of such $n$ is twice the dimension of the representation. In the following 
subsections we will give the explicit form of these representations with a sketch of 
the methods used to derive them. 

The decomposition of any representation of \su\ in irreducible representations of 
$\Gamma$ is therefore reduced to a combinatorial problem of multiplying polynomials to 
get objects of the desired degrees. The linear independence of the resulting 
realizations can be checked.  The  normalizations of the obtained 
polynomials cannot be predicted from the one of their components, and we therefore 
just try to have the simplest coefficients. As it should be, vectors belonging to 
distinct irreducible representations are  automatically orthogonal. However, as soon 
as a given representation can be obtained in different ways at a 
given degree, we do not generally obtain orthogonal elements. In applications, it is 
generally simpler computationally not to build explicitly an orthonormal basis but to 
use the Gran--Schmidt matrix of scalar product to obtain basis independent results. 
Interestingly, these Gran--Schmidt matrices have rather peculiar 
arithmetic properties which we develop in the appendix~\ref{norm}.

In the case of the cyclic group of order $n$ and the dihedral group of order $4n$, 
the corresponding objects can be trivially written. The invariants for the cyclic group 
are $st$ and $s^{n}+t^{n}$. All representations are of dimension 1 and appear two 
times, the trivial one with $1$ and $s^{n}-t^{n}$, the $n-1$ others 
as $s^{i}$ and $t^{n-i}$, with $i$ varying between 1 and $n-1$. For the dihedral 
groups, 
the invariants are $s^{2}t^{2}$ and $s^{2n}+t^{2n}$. The trivial representation 
appears as $1$ and $st(s^{2n}-t^{2n})$. A second one-dimensional representations appears 
with $st$ and $s^{2n}-t^{2n}$.
The two-dimensional representations have four realizations with $i$ also varying 
between 1 and $n-1$:
the doublets $(s^{i},t^{i})$, $(s^{i+1}t, - s t ^{i+1})$, $(t^{2n-i},(-1)^{i}s^{2n-i})$ and $(st^{2n-i+1},(-1)^{i+1}s^{2n-i+1}t)$. The last two representations depend on 
the parity of $n$, since they are bosonic for $n$ even and fermionic otherwise. For 
$n$ even, we have one representation given by $s^{n}+t^{n}$ or 
$st(s^{n}-t^{n})$ and the other given by $s^{n}-t^{n}$ or $st(s^{n}+t^{n})$. For $n$ 
odd, the imaginary unit $i$ appears. We therefore have $s^{n}+i t^{n}$ or $st(s^{n}-i 
t^{n})$ for one representation and $s^{n}-i t^{n}$ or $st(s^{n}+i t^{n})$ for the 
other.

\subsection{Tetrahedron and octahedron.}
The tetrahedral group is generated by $i$, $j$ and $1/2(1+i+j+k)$. In degree~4, 
$s^{4}
+t^{4}$ and $s^{2}t^{2}$ are invariant under $i$ and $j$ and we can form with them 
covariants of the group, that is objects transforming under the non-trivial 
one-dimensional representations of the tetrahedral group $T= s^{4}+t^{4}\pm2i\sqrt{3} 
s^{2}t^{2}$. From them every representations can be formed.

\begin{table}[t]
\caption{Representations of the binary tetrahedral group.}
\begin{center}
$\begin{array}{|c|c|}
\hline1 & (1) \\
&(t^{12}-33\,s^4\,t^8-33\,s^8\,t^4+s^{12})\\
\hline
& (s,t)\\
2&(s^{5}-5 \,s \,t^{4},t^{5}-5\,s^{4}\,t)\\
&(-7\,s^{4}\,t^{3}-t^{7},s^{7}+7\,s^{3}\,t^{4})\\
&(11\,s^{8}\,t^{3}+22\,s^{4}\,t^{7}-t^{11},s^{11}-22\,s^{7}\,t^{4}-11 \,s^{3}\,t^{8})\\[2mm]
\hline
&(s^{2},s\,t,t^{2})\\
&(4 \,s \,t^{3},s^{4}-t^{4},-4\,s^{3}\,t)\\ 
3&(s^{6}-5\,s^{2}\,t^{4},-2(s^{5}\,t + s\, t^{5}),t^{6}-5 \,s^{4}\,t^{2})\\
&(t^{6}+3\,s^{4}\,t^{2},-4 \,s^{3}\,t^{3},s^{6}+3\,s^{2}\,t^{4})\\
&(-2\,s\,t^{7}-14\,s^{5}\,t^{3},s^{8}-t^{8},2\,s^{7}\,t+14\,s^{3}\,t^{5})\\
&( t^{10}-14\,s^4\,t^6-3\,s^8\,t^2,8\,s^3\,t^7+8\,s^7\,t^3,-3\,s^2\,t^8-14\,s^6\,t^4+s^{10}) \\
\hline
&(-t^{3}\mp i\,\sqrt3\, s^{2}\,t,s^{3}\pm i\,\sqrt3 \,s \,t^{2})\\
2_{1,2}&(s^5+s\,t^4\pm2\,i\,\sqrt{3}\,s^3\,t^2,t^5+s^4\,t\pm2\,i\,\sqrt{3}\,s^2 
\,t^3)\\
&\bigl(-t^7+5\,s^4\,t^3\pm i\,\sqrt3\,(s^6\,t+3\,s^2\,t^5),s^7-5\,s^3\,t^4\mp 
i\,\sqrt3\,(s\,t^6+3\,s^5\,t^2)\bigr)\\
&\scriptstyle\bigl(s\,t^8-10\,s^5\,t^4+s^9\mp4\, i\,\sqrt3\,(s^3\,t^6+s^7\,t^2),
 t^9-10\,s^4\,t^5+s^8\,t\mp4\, i\,\sqrt3\,(s^2\,t^7+s^6\,t^3)\bigr)\\
\hline
1_{1,2}&(s^{4}+t^4\pm2\,i\,\sqrt{3}\,s^2\,t^2)\\
&\bigl(t^8-10\,s^4\,t^4+s^8\mp4\,i\,\sqrt3(s^{6}\,t^{2}+s^{2}\,t^{6})\bigr)\\ 
\hline
\end{array}$
\end{center}
\label{tetra}
\end{table}%

The squares give the two other required instances of the one-dimensional 
representations. Acting on them with the partial derivatives $-\partial/\partial t$ and 
$\partial/\partial s$, or multiplying them with $s$ and $t$ yields the four instances 
of the $2_{1}$ and $2_{2}$ representations.  
Their product gives the invariant of  degree~8, $P=s^{8} +14 s^{4}t^{4} + t^{8} $ and 
their combination with the cross product $C$ gives the invariant of degree~6, $Q= s 
t^{5} - s^{5} t$. The cross product of these two invariants gives the invariant of 
degree~12, $s^{12} - 33 s^{8} t^{4} - 33 s^{4} t^{8} + t^{12}$, 
which can also be obtained from the real part of the third power of $T$, the 
imaginary part being proportional to $Q^{2}$.
The fundamental representation is obtained by acting with the derivatives on the 
three invariants of degree~6, 8 and~12.  The vector representation can be obtained from 
the cross product of the basic one
$(s^{2},st,t^{2})$ with the two invariants P and Q, and also by applying the triplet 
$(\partial^{2}/\partial s^{2},-\partial^{2}/\partial s\partial t, 
\partial^{2}/\partial s^{2})$ on the three invariants.

It is therefore possible to construct the full decomposition of the representations 
of \su\ under the tetrahedral group without having to make explicit calculations of 
the transformations, apart from the covariance property of $T$. The necessary 
polynomials are collected in table~(\ref{tetra}). The labeling of the 
representations is given on the extended E6 diagram, following the McKay 
correspondence.
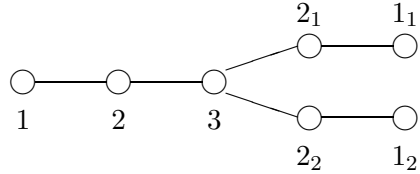
\[
\setlength{\unitlength}{3947sp}%
\begin{picture}(2713,1114)(68,-269)
{\thinlines
\put(751,314){\circle{150}}
}%
{\put(1351,314){\circle{150}}
}%
{\put(1951,539){\circle{150}}
}%
{\put(1951, 89){\circle{150}}
}%
{\put(2551, 89){\circle{150}}
}%
{\put(2551,539){\circle{150}}
}%
{\put(151,314){\circle{150}}
}%
{\put(226,314){\line( 1, 0){450}}
}%
{\put(1426,389){\line( 3, 1){450}}
}%
{\put(1426,239){\line( 3,-1){450}}
}%
{\put(826,314){\line( 1, 0){450}}
}%
{\put(2026,539){\line( 1, 0){450}}
}%
{\put(2026, 89){\line( 1, 0){450}}
}%
\put( 151, 14){\makebox(0,0)[b]{\smash{{$1$}%
}}}
\put(751, 14){\makebox(0,0)[b]{\smash{{$2$}%
}}}
\put(1351, 14){\makebox(0,0)[b]{\smash{{$3$}%
}}}
\put(2551,-211){\makebox(0,0)[b]{\smash{{$1_2$}%
}}}
\put(1951,-211){\makebox(0,0)[b]{\smash{{$2_2$}%
}}}
\put(1951,689){\makebox(0,0)[b]{\smash{{$2_1$}%
}}}
\put(2551,689){\makebox(0,0)[b]{\smash{{$1_1$}%
}}}
\end{picture}%
\]

The corresponding decomposition for the octahedral group is rather simple now, since 
the tetrahedral group is a subgroup of index two. The representations $1_{1}$ and 
$1_{2}$ of the tetrahedral group merge in a single two dimensional representation 
$2_{V}$, the representations $2_{1}$ and $2_{2}$ give the four 
dimensional representation. The 1, 2 and 3 representations of the tetrahedral group 
each appear in two different versions as representations of the octahedral group, 
which can be interchanged by any object in the $1'$ representation. The relation 
between representations of the binary tetrahedral group and the binary 
octahedral group appear in the following unusual disposition of the extended E7 
Dynkin diagram.
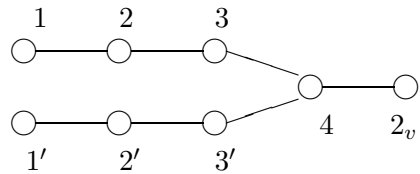
\[
\setlength{\unitlength}{3947sp}%
\begin{picture}(2641,1081)(668,-260)
{\thinlines
\put(3226,314){\circle{150}}
}%
{\put(2626,314){\circle{150}}
}%
{\put(2026,539){\circle{150}}
}%
{\put(2026, 89){\circle{150}}
}%
{\put(1426, 89){\circle{150}}
}%
{\put(1426,539){\circle{150}}
}%
{\put(826,539){\circle{150}}
}%
{\put(826, 89){\circle{150}}
}%
{\put(2551,389){\line(-3, 1){450}}
}%
{\put(2551,239){\line(-3,-1){450}}
}%
{\put(3151,314){\line(-1, 0){450}}
}%
{\put(1951,539){\line(-1, 0){450}}
}%
{\put(1951, 89){\line(-1, 0){450}}
}%
{\put(1351,539){\line(-1, 0){450}}
}%
{\put(1351, 89){\line(-1, 0){450}}
}%
\put(976,689){\makebox(0,0)[rb]{\smash{{{$1$}%
}}}}
\put(2776, 14){\makebox(0,0)[rb]{\smash{{{$4$}%
}}}}
\put(3301, 14){\makebox(0,0)[rb]{\smash{{{$2_v$}%
}}}}
\put(1426,689){\makebox(0,0)[lb]{\smash{{{$2$}%
}}}}
\put(2026,689){\makebox(0,0)[lb]{\smash{{{$3$}%
}}}}
\put(2026,-211){\makebox(0,0)[lb]{\smash{{{$3'$}%
}}}}
\put(1426,-211){\makebox(0,0)[lb]{\smash{{{$2'$}%
}}}}
\put(976,-211){\makebox(0,0)[rb]{\smash{{{$1'$}%
}}}}
\end{picture}%
\]
A simple multiplication by the covariant $Q$, which transform in the $1'$ 
representation of the octahedral group allows to obtain all necessary construct for the 
decomposition under this group. We will not write them explicitly.

\subsection{Dodecahedral group}
The corresponding calculations for the dodecahedral group are more complex, since the 
representations are bigger and have more different avatars. In order to limit the 
place taken, I will only give explicitly the vectorial representations, which are the 
only necessary ones for our project, apart from the alternative 
two dimensional one, which is useful to build the vectorial four dimensional 
representation.

As before we need to make a specific choice for the orientation of the polyhedron. It 
will be very convenient to follow Klein in choosing to orient a five fold symmetry 
axis along the $z$-axis. The orientation is completely fixed by choosing to make a 
two fold symmetry axis coincide with the $y$-axis. This single out a 
ten element dihedral subgroup of the dodecahedral group or rather its twenty elements 
double cover as a subgroup of the binary dodecahedral group. One particularly useful 
property of this orientation is that, apart in one case, the eigenvector spaces of 
the order five rotation are one-dimensional in the 
irreducible representations of the dodecahedral group, the exception being the one 
corresponding to the eigenvalue $-1$ in the six-dimensional spinorial representation. 
This allows for an easy link between the vectors of different realizations of the 
same irreducible representation.
The full group can then be generated from this dihedral subgroup and a single 
supplementary element, since no subgroup of the dodecahedral group has index 2 or~3. Our 
choice is for a two fold symmetry with an axis in the $xz$ plane. Apart from the 
normalization factor, which is never very important, it can be chosen 
with rather simple coordinates:
\begin{equation}\label{special}
K'={1+\sqrt5\over 2}k - i
\end{equation}

According to the McKay correspondence, the irreducible representations of the 
dodecahedral group label the vertex of the extended E8 diagram:
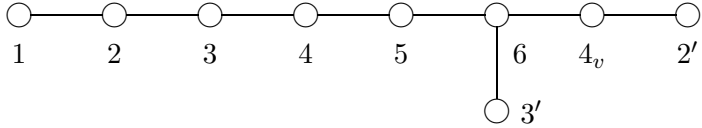
\[
\setlength{\unitlength}{3947sp}%
\begin{picture}(4634,807)(699,-185)
{\thinlines
\put(2026,539){\circle{150}}
}%
{\put(1426,539){\circle{150}}
}%
{\put(826,539){\circle{150}}
}%
{\put(3826,539){\circle{150}}
}%
{\put(3226,539){\circle{150}}
}%
{\put(2626,539){\circle{150}}
}%
{\put(5026,539){\circle{150}}
}%
{\put(4426,539){\circle{150}}
}%
{\put(3826,-61){\circle{150}}
}%
{\put(1951,539){\line(-1, 0){450}}
}%
{\put(1351,539){\line(-1, 0){450}}
}%
{\put(2551,539){\line(-1, 0){450}}
}%
{\put(4351,539){\line(-1, 0){450}}
}%
{\put(3751,539){\line(-1, 0){450}}
}%
{\put(3151,539){\line(-1, 0){450}}
}%
{\put(4951,539){\line(-1, 0){450}}
}%
{\put(3826, 14){\line( 0, 1){450}}
}%
\put(5026,239){\makebox(0,0)[b]{\smash{{{$2'$}%
}}}}
\put(826,239){\makebox(0,0)[b]{\smash{{{$1$}%
}}}}
\put(1426,239){\makebox(0,0)[b]{\smash{{{$2$}%
}}}}
\put(2026,239){\makebox(0,0)[b]{\smash{{{$3$}%
}}}}
\put(2626,239){\makebox(0,0)[b]{\smash{{{$4$}%
}}}}
\put(3226,239){\makebox(0,0)[b]{\smash{{{$5$}%
}}}}
\put(3926,239){\makebox(0,0)[l]{\smash{{{$6$}%
}}}}
\put(4426,239){\makebox(0,0)[b]{\smash{{{$4_v$}%
}}}}
\put(3976,-136){\makebox(0,0)[lb]{\smash{{{$3'$}%
}}}}
\end{picture}%
\]
The first step is to determine the lowest dimensional non trivial splitting, which 
occurs for the homogeneous polynomials of degree 6, the representation $l=3$ of \su. 
This is made easy because $s^{3}t^{3}$ belongs to the $3'$ representation and the 
doublet $(s^{4}t^{2},s^{2}t^{4})$ belongs to the vectorial $4_{v}$ 
representation. Use of the transformation in~(\ref{special}) then allows to complete 
the representations.  The $3'$ has a basis given by
\begin{equation}
 (3\,s\,t^5+s^6,5\,s^3\,t^3,t^6-3\,s^5\,t).
\end{equation}
The basis of the $4_{v}$ representation is 
\begin{equation}
(-t^6-2\,s^5\,t,5\,s^4\,t^2,5\,s^2\,t^4,s^6-2\,s\,t^5).
\end{equation}
The factors of 5 are chosen to simplify the transformations and to have simple 
relations between the norms of the basic vectors. As it should be, vectors belonging to 
different representations are orthogonal. The fact that the $4_{v}$ representation is 
the tensor product of the basic one and the second two dimensional 
one $2'$ allows to determine the basis of the $2'$ representation with polynomials of 
degree~7. They should reproduce the basis elements of the degree~6 representation 
$4_{v} $ upon differentiation with respect to $s$ and $t$. One obtains
\begin{equation}
(t^{7}+7\,s^{5}\,t^{2},s^{7}-7\,s^{2}\,t^{5})
\end{equation}
\begin{sidewaystable}
\caption{Representations of the binary dodecahedral group.\label{dotable}}
\begin{center}
$\begin{array}{|c|c|}
\hline 1 & (1) \\
&(t^{30}-522\,s^5\,t^{25}-10005\,s^{10}\,t^{20}-10005\,s^{20}\,t^{10}+522\,s^{25}\,t^5+s^{30})\\
\hline
&(s^{2},s\,t,t^{2})\\
&(30\,s^6\,t^4-10\,s\,t^9,t^{10}-36\,s^5\,t^5-s^{10},30\,s^4\,t^6+10\,s^9\,t)\\
3&(-11\,s^2\,t^{10}+66\,s^7\,t^5+s^{12},-5\,s\,t^{11}-5\,s^{11}
 \,t,t^{12}-66\,s^5\,t^7-11\,s^{10}\,t^2)\\
&(t^{18}+126\,s^5\,t^{13}+117\,s^{10}\,t^8-12\,s^{15}\,t^3,
-45\,s^4\,t^{14}-130\,s^9\,t^9+45\,s^{14}\,t^4,
12\,s^3\,t^{15}+117\,s^8 \,t^{10}-126\,s^{13}\,t^5+s^{18}) \\
&(2\,s\,t^{19}+342\,s^6\,t^{14}+494\,s^{11}\,t^9-114\,s^{16}\,t^4,
t^{20}+114\,s^5\,t^{15}+114\,s^{15}\,t^5-s^{20},
-114\,s^4\,t^{16}-494\,s^9\,t^{11}+342\,s^{14}\,t^6-2\,s^{19}\,t)\\
&(t^{28}-360\,s^5\,t^{23}-4370\,s^{10}\,t^{18}-1035\,s^{20}\,t^8+12\,s^{25}\,t^3,
75\,s^4\,t^{24}+2300\,s^9\,t^{19}+2300\,s^{19}\,t^9-75\,s^{24}\,t^4,
\ldots)\\
\hline
& (s^4,2\,s^3\,t,3\,s^2\,t^2,2\,s\,t^3,t^4)\\
&(2\,s^6\,t^2-4\,s\,t^7,t^8-8\,s^5\,t^3,15\,s^4\,t^4,-8\,s^3\, t^5-s^8,2\,s^2\,t^6+4\,s^7\,t)\\
&(6\,s^2\,t^8-8\,s^7\,t^3,2\,s\,t^9+14\,s^6\,t^4,-t^{10}-s^{10},
2\,s^9\,t-14\,s^4\,t^6,8\,s^3\,t^7+6\,s^8\,t^2)\\
&\bigl(10\,s^3\,t^3\,(t^6-3\,s^5\,t),(3\,s\,t^5+s^6)^2,
 3\,s\,t^{11}+42\,s^6\,t^6-3\,s^{11}\,t,-(t^6-3\,s^5\,t)^2,
 -10\,s^3\,t^3\,(3\,s\,t^5+s^6)\bigr)\\
5&(-11\,s^4\,t^{10}+66\,s^9\,t^5+s^{14},
-16\,s^3\,t^{11}+66\,s^8\,t^6-4\,s^{13}\,t,-15\,s^2\,t^{12}-15\,s^{12}\,t^2,
-4\,s\,t^{13}- 66\,s^6\,t^8-16\,s^{11}\,t^3,t^{14}-66\,s^5\,t^9-11\,s^{10}\,t^4)\\
 &(17\,t^{16}+1092\,s^5\,t^{11}+364\,s^{10}\,t^6-4\,s^{15}\,t,
 -910\,s^4\,t^{12}-1040\,s^9\,t^7+60\,s^{14}\,t^2,
 420\,s^3\,t^{13}+1755\,s^8\,t^8-420\,s^{13}\,t^3,
\ldots) \\
&(-2\,s\,t^{17}-182\,s^6\,t^{12}-104\,s^{11}\,t^7+4\,s^{16}\,t^2,
 -t^{18}+14\,s^5\,t^{13}+143\,s^{10}\,t^8-28\,s^{15}\,t^3,
 75\,s^4\,t^{14}+75\,s^{14}\,t^4,
\ldots)\\
&\bigl((30\,s^6\,t^4-10\,s\,t^9)^2,2\,(30\,s^6\,t^4-10\,s\,t^9)\,(t^{10}-36\,s^5\,t^5-s^{10}),
2\,t^{20}-444\,s^5\,t^{15}+3388\,s^{10}\,t^{10}+444\,s^{15}\,t^5+2\,s^{20},
\ldots \bigr)\\
&(-2\,s^3\,t^{19}-342\,s^8\,t^{14}-494\,s^{13}\,t^9+114\,s^{18}\,t^4,
-3\,s^2\,t^{20}-456\,s^7\,t^{15}-494\,s^{12}\,t^{10}+s^{22},
-3\,s\,t^{21}-342\,s^6\,t^{16}-342\,s^{16}\,t^6+3\,s^{21}\,t,\ldots) \\
& {\scriptstyle(21\,t^{26}-5060\,s^5\,t^{21}-37145\,s^{10}\,t^{16}-1610\,s^{20}\,t^6+2\,s^{25}\,t, 2300\,s^4\,t^{22}+43700\,s^9\,t^{17}+9200\,s^{19}\,t^7-50\,s^{24}\,t^2,
-600\,s^3\,t^{23}-32775\,s^8\,t^{18}-32775\,s^{18}\,t^8+600\,s^{23}\,t^3,\ldots)}\\
\hline
&(3\,s\,t^5+s^6,5\,s^3\,t^3,t^6-3\,s^5\,t)\\
&(10\,s^8\,t^2-20\,s^3\,t^7,t^{10}+14\,s^5\,t^5-s^{10},10\,s^2 \,t^8+20\,s^7\,t^3)\\
3'&(t^{14}+14\,s^5\,t^9+49\,s^{10}\,t^4,-7\,s^2\,t^{12}-48\,s^7 \,t^7+7\,s^{12}\,t^2,
49\,s^4\,t^{10}-14\,s^9\,t^5+s^{14})\\
&(3\,s\,t^{15}+143\,s^6\,t^{10}-39\,s^{11}\,t^5-s^{16},25\,s^3\,t^{13}+25\,s^{13}\,t^3,
-t^{16}+39\,s^5\,t^{11}+143\,s^{10}\,t^6-3\,s^{15}\,t)\\
&(-52\,s^3\,t^{17}-442\,s^8\,t^{12}-544\,s^{13}\,t^7+14\,s^{18}\,t^2,
-t^{20}+136\,s^5\,t^{15}+136\,s^{15}\,t^5+s^{20},
14\,s^2\,t^{18}+544\,s^7\,t^{13}-442\,s^{12}\,t^8+52\,s^{17}\,t^3)\\
&(-t^{24}+112\,s^5\,t^{19}+646\,s^{10}\,t^{14}-1292\,s^{15}\,t^9+119\,s^{20}\,t^4,
5\,s^{22}\,t^2-5\,s^2\,t^{22}-570\,(s^7\,t^{17}+s^{17}\,t^7),
\ldots)\\
\hline
&(t^7+7\,s^5\,t^2,s^7-7\,s^2\,t^5)\\
2'&(-26\,s^3\,t^{10}-39\,s^8\,t^5+s^{13},-t^{13}-39\,s^5\,t^8+26\,s^{10}\,t^3)\\
&(-t^{17}+119\,s^5\,t^{12}-187\,s^{10}\,t^7+17\,s^{15}\,t^2,
-17\,s^2\,t^{15}-187\,s^7\,t^{10}-119\,s^{12}\,t^5-s^{17})\\
&(-46\,s^3\,t^{20}-1173\,s^8\,t^{15}+391\,s^{13}\,t^{10}-207\, s^{18}\,t^5-s^{23},
 t^{23}-207\,s^5\,t^{18}-391\,s^{10}\,t^{13}-1173\,s^{15}\,t^8+46\,s^{20}\,t^3)\\ 
\hline
\end{array}$
\end{center}
\label{default}
\end{sidewaystable}%
A representation $4_{v}$ in degree 8 is obtained by multiplying this vector by $s$ 
and $t$.
From the $3'$ representation, we can form an invariant norm, which is the invariant 
polynomial of degree~12, $P=s\,t^{11}-11\,s^{6}\,t^{6}-s^{11}\,t$. The Hessian gives 
then the second invariant of degree~20,
$$t^{20}+228\,s^5\,t^{15}+494\,s^{10}\,t^{10}-228\,s^{15}\,t^5+s^{20}$$
The three  antisymmetric combinations of the basis functions of the first $3'$ 
representations by the operation of~(\ref{combeq}) form a new representations $3'$ in 
degree~10. A third version of the $3'$ representation appears as the symmetric square 
of the $2'$ representation, the one of degree~14. The three other 
versions of the $3'$ representation can be obtained by combination with the 
invariant~$P$, adding 10 to the degree. The $2'$ representation in degree~13 can be obtained 
by combining the $3'$ representation in degree 6 and the $2'$ representation in 
degree~7, but it is simpler to use the combination method of the 
appendix~\ref{comb}. This is also handy to produce all the occurrences of the $3$ and 
$5$ representations from their fundamental case and the invariants. Finally, the 
useful representations are summarized in table~\ref{dotable}. For lack of space, some 
of the representations are not complete. The other basis elements 
can be obtained from the written ones by using the rotation around the $y$ axis, 
which stems from the simple substitution $s \to t$ and $ t \to -s$. The $4_{v}$ and the 
$6$ representations can be obtained from the $2'$ and $3'$ representations, by using 
the two possible ways of 
tensoring by the fundamental representation, that is multiplying each element either 
by $s$ or $t$, or differentiating by $-\pa_{t}$ and $\pa_{s}$.

A method for the decomposition of the representation of \su\ in irreducible 
representations of the binary icosahedral group was proposed in~\cite{Kra05}. With respect to 
the present work, it presents however strong limitations, since it is based on 
diagonalizing an operator. Explicit solutions become difficult to 
obtain as soon as the blocks to diagonalize have a rank bigger than 2 or 3. 
Furthermore, since irreducible representations are recognized as the eigenspaces of the 
operator, they do not come with a standard basis in which the representation of the 
group $\Gamma$ is fixed. This does not allow to use the symmetry 
under $\Gamma$ to relate the correlations for different vectors of the same 
representation.

\section{Conclusion}
An explicit decomposition of the representations of the group \su\ in irreducible 
representations of its finite subgroups has been provided.  This allows not only to 
produce all eigenmodes of the quotients of the three-sphere, but also to group them 
in irreducible representations of the group $\Gamma$. In this way, the 
determination of the correlation matrix of the different modes of the CMB becomes 
accessible. The kind of analysis which has only be made for toroidal geometries, 
allowing for a maximal use of the available evidence to differentiate the possible 
geometries, can now be done also for spherical 
geometries.

{\bf Acknowledgments:} It is a pleasure to thank Olivier Babelon and Jean-Bernard 
Zuber who gave me access to the references~\cite{klein,Ko}. The exigence of an anonymous 
referee contributed to the clarity of this work.

\appendix
\section{Representation of \su\ as homogeneous polynomials.} \label{comb}
It is a well known fact that the irreducible representations of \su\ can all be 
obtained as symmetrical tensor products of the fundamental one. This has motivated the 
expression of the components of any representation as multispinors and has been 
extensively used to deduce properties of the representation matrix and 
the Clebsch--Gordan coefficients for the decomposition of tensor products of 
representations, see e.g., \cite{LanLif}. But polynomials are just that, the symmetric 
product of objects which can be considered as basis elements in a fundamental 
representation.
They are very interesting for automated treatments since most computer algebra 
systems have in their kernel of functions the operations on polynomials.

The action of the group \su\ on any object is obtained through substitution for the 
basic variables of their linear transform. This is quite easy. What is not so 
straightforward is the expression of the decomposition of tensor product of 
representations in their irreducible factors. Let us consider homogeneous 
polynomials $P(s,t)$ of degree $p$ and $Q(s,t)$ of degree $q\leq p$, representing the 
objects of spin $p/2$ and $q/2$ for which we wish the decomposition in irreducible 
components.  The highest spin component is easily obtained, since the product $PQ$ 
gives a polynomial with degree $p+q$. The lowest spin component 
can also be inferred in substituting in $Q$ the derivatives $-\partial_{t}$ and 
$\partial_{s}$ to the variables $s$ and $t$: 
$Q(-\partial_{t},\partial{s}) P(s,t)$ is a polynomial of degree $p-q$.

In the general case, the idea is to enlarge the representation by a factor which will 
be exactly compensated by a similar one in the other one. This role is played by 
differential operators of order $r$ acting on $P$ and $Q$ combined to form an 
invariant of \su.  The representation of degree $p+q-2r$ is therefore 
obtained through
\begin{equation} \label{combfor}
\sum_{k=0}^{r} (-1)^{k} \pmatrix{r\cr k\cr} {\partial^{r}\over \partial s^{k}\partial 
t^{r-k} }P(s,t)
	{\partial^{r}\over \partial s^{r-k}\partial t^{k} }Q(s,t)
\end{equation}
Apart from a normalization factor which depends on $p$, $q$ and $r$, this formula can 
be shown to be equivalent to the usual formulas for $3j$-symbols. This normalization 
could be obtained recursively, taking for $P$ the monomial $s^{p}$ and for $Q$ the 
monomials $s^{q-l}t^{l}$ for successive values of $l$, but I shall 
not need it in this work. The primary representations which we obtain do not need 
normalization and only relative normalization will be used in the study of the 
asymmetries of the microwave background. The use of the non-normalized basis of the 
homogeneous polynomials has the advantage of ridding ourselves 
of all square roots factors. This is computationally interesting since the extraction 
of the squared factors under square roots is much more demanding that purely 
rational calculations.

\section{Remarks on the norms.}\label{norm}
It is quite remarkable that the squared norm of the objects created in 
section~(\ref{McK}) are peculiar rational numbers. The numerator is, apart from small eventual 
powers of 2 and 3, a power of 5 and the denominator is the product of primes smaller 
than the degree of the object. This pattern seems to fail for bigger 
degrees, but it can be observed that the failure happens when there are independent 
objects of the same degrees with the same transformation properties. In this case 
there is an arbitrariness in the choice of the basis functions, so that it is natural 
that the objects obtained by arbitrary choices do not 
possess any particular property. 

An intrinsic description of the space of the invariants can be obtained with a 
projector. It does not depend on any base choice and the matrix elements of the projector 
are then rational numbers with denominators which are essentially powers of 5, i.e., 
eventually multiplied by 3, 2 or 4. In the case of the tetrahedron 
group or the octahedron group, the same pattern appears with 5 replaced by 2. 

This suppose that we express the projector in the basis of the homogeneous 
polynomials, which is not normalized. Otherwise we obtain a symmetrized matrix with square 
roots appearing in the non-diagonal elements.

This pattern can be understood since the projector on the space of invariants can be 
obtained as the mean of the group elements in the group algebra. Then we can use the 
explicit expression given by Klein for the elements of the group which are not in the 
binary dihedral group of order 20, in term of a fifth root of 
the unity $\epsilon$.
\begin{eqnarray}
\sqrt{5} \pmatrix{ s' \cr t'\cr} &=& \pm \pmatrix{\epsilon&0\cr 0&\epsilon^{4}}^{\nu}
\pmatrix{ - \epsilon+\epsilon^{4} & \epsilon^{2}-\epsilon^{3}\cr \epsilon^{2}-\epsilon^{3}& 
\epsilon-\epsilon^{4}\cr}  \pmatrix{\epsilon&0\cr 0&\epsilon^{4}}^{\mu} \pmatrix{s 
\cr t\cr} \label{kl1}\\
\sqrt{5} \pmatrix{ s' \cr t'\cr} &=& \pm \pmatrix{\epsilon&0\cr 0&\epsilon^{4}}^{\nu}
\pmatrix{ - \epsilon^{2}+\epsilon^{3} &- \epsilon+\epsilon^{4}\cr -\epsilon+\epsilon^{4}& 
\epsilon^{2}-\epsilon^{3}\cr}  \pmatrix{\epsilon&0\cr 0&\epsilon^{4}}^{\mu} 
\pmatrix{s \cr t\cr}
\label{kl2}\end{eqnarray}
The matrix notations are not those of Klein and the exponents $\mu$ and $\nu$ take 5 
different values. When we take for $\epsilon$ a different fifth root of the unity, 
the elements described by eqs.~(\ref{kl1},\ref{kl2}) are permuted among themselves. 
The sum on the elements of the group is therefore independent on the 
choice of the root $\epsilon$, that is an invariant of the Galois group and therefore 
a rational number. Furthermore, $\epsilon$ is an integral element of its field, 
since it is a solution of an equation with integer coefficients with leading 
coefficient 1. When we apply the transformations 
(\ref{kl1},\ref{kl2}) to homogeneous polynomials of degree $2n$, the $\sqrt 5$ factor 
gives a $5^{n}$ denominator and the numerator is polynomial in $\epsilon$, $s$ and 
$t$ with integer coefficients, so that when we sum over the group elements we obtain 
a polynomial in $s$ and $t$ with integer coefficients for the 
numerator. In the basis of the monomials in $s$ and $t$, the projection on the 
invariants is therefore given by a matrix with integer coefficients divided by $60.5^{n}$. 
The factor of 60 stems from the division by the order of the group.

In the case of non-trivial representations of the group $\Gamma$, the projections on 
specific basis elements of the representation as well as the elementary matrix 
sending from one basis element to an other should be expressible as elements of the 
group algebra and a similar result should be proved. However the 
coefficients of these elements of the group algebra are no longer rational and the 
invariance of the resulting matrices under the Galois group depends on a subtle 
relation between the exchange of coefficients and of group elements under the Galois 
group. The cases of the tetrahedron or octahedron groups are 
similar.

\end{document}